# X-ray/Optical Variability of Akn 120 and 3C 120

K. Marshall*, E. C. Ferrara†, H. R. Miller*, A. P. Marscher** and G. Madejski‡

*Dept. of Physics & Astronomy, Georgia State University, Atlanta, GA 30303
†General Dynamics, 7915 Canter Court, Severn, MD 21144
**Department of Astronomy, Boston University, 725 Commonwealth Avenue, Boston, MA 02215
‡Stanford Linear Accelerator Center, Glast Group, 2575 Sand Hill Road, MS 43A, Menlo Park, CA 94025

**Abstract.** Using the Rossi X-ray Timing Experiment (RXTE), we have observed both the Seyfert 1 galaxy Akn 120 and the broad-line radio galaxy 3C 120. Monitoring observations spanning 2 years plus intensive stares spanning several days have been obtained for both objects. These data have allowed us to construct the power density spectrum (PDS) for each object. Covering more than 4 decades in temporal frequency, the PDS for 3C 120 represents the first such measurement for a radio loud galaxy. The PDS for 3C 120 shows a steep power law at high frequencies, which turns over at low frequencies. Akn 120 also has a power law shape in its PDS. Using the PDS of Cygnus X-1, we estimate the masses for both of these objects. Optical data and cross correlation functions are also presented for Akn 120. The long-term light curve of Akn 120 shows a strong correlation with optical data, with zero time lag.

## INTRODUCTION

In recent years, the power density spectrum (PDS) has manifested itself as one of a growing number of methods for calculating the mass of the central compact object in active galactic nuclei (AGN) [1, 2, 3]. This technique has been applied to several radio-quiet objects, with positive results. In this contribution, we present the PDSs for 3C 120 and Akn 120. The PDS for 3C 120 represents the first such measurement for a radio-loud galaxy. We also present results of an extensive optical monitoring campaign for Akn 120.

## OBSERVATIONS AND DATA REDUCTION

Constructing a PDS requires observations spanning a wide range of timescales. For an AGN, this can mean up to 4 orders of magnitude in temporal frequency, in order to accurately constrain the shape and possible break frequency of the PDS. The flexible observing schedule of RXTE allows us to easily observe 3C 120 and Akn 120 on multiple timescales, with very even sampling.

Marscher et al. have carried out an extensive monitoring program, observing 3C 120 once per week for all of 1998, followed by twice per week for 1999 [4]. These data are very evenly sampled, with few missing points aside from an 8-week gap due to 3C 120's proximity to the sun. We also use data from observations taken every 2 days for $\sim 10$ months, part of an ongoing monitoring program by Marscher.

Data have also been obtained from a continuous stare, covering approximately 9 days during 18-27 Dec. 2002. Results from this observation were re-binned at 1000s. The data contain some gaps due to South Atlantic Anomaly (SAA) passage and interruptions for other observing programs. These gaps were interpolated linearly between the data points just before and just after the gaps. Despite a large number of gaps ($\sim 60\%$ data missing) for this observation, the light curve presented in Figure 1 shows that the variability on short timescales is clearly resolved. Parameters for all our observations of 3C 120 are presented in Table 1.

Similarly for Akn 120, we have obtained 2 years of monitoring observations taken every 3 days. A 3-day intensive stare provides data on short timescales. Again, the short timescale light curve shows a significant number of gaps. Table 2 shows parameters for our observation of Akn 120.

Light curves were extracted from all data sets using the FTOOLS v5.2 software package[1]. During nearly all of our observations PCUs 3 and 4 were turned off. Also, for our most recent data PCU1 was turned off. The loss of the PCU0 propane layer in May 2000 resulted in added complications for background modeling, therefore we chose only to extract data from Layer 1 of PCU2. We extracted data using the standard extraction criteria, using channels 0–27 (2–11 keV). The faint-mode "L7" model was used to provide background estimation. Figure 1 and Figure 2 show light curves for 3C 120 and Akn 120, respectively.

---

[1] Provided by HEASARC

**TABLE 1.** Sampling parameters for 3C 120 data.

| Timescale | MJD Dates | Sampling Interval | $N^*$ | Percent Missing[†] | $\mu^{**}$ |
|---|---|---|---|---|---|
| Long | 50812.06–51563.18 | 7 d | 107 | 18.69 | 4.66 |
| Intermediate | 52450.08–52756.28 | 2 d | 155 | 2.86 | 5.31 |
| Short | 52622.40–52631.22 | 1000 s | 786 | 60.05 | 5.70 |

[*] Total number of data points.
[†] Percentage of data points which require interpolation.
[**] Mean count rate.

**TABLE 2.** Sampling parameters for Akn 120 data.

| Timescale | MJD Dates | Sampling Interval | $N^*$ | Percent Missing[†] | $\mu^{**}$ |
|---|---|---|---|---|---|
| Long | 50868.09–51254.29 | 3 d | 174 | 34.58 | 3.55 |
| Short | 51162.70–51165.52 | 2000 s | 122 | 41.80 | 3.62 |

[*] Total number of data points.
[†] Percentage of data points which require interpolation.
[**] Mean count rate.

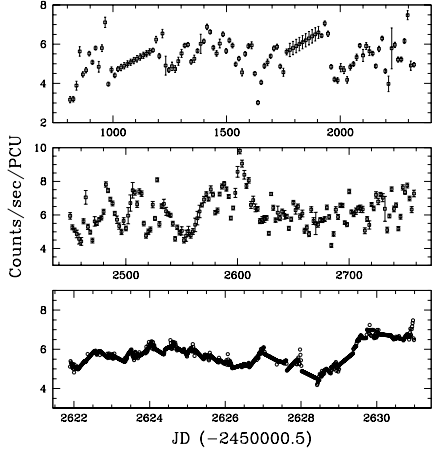

**FIGURE 1.** Light curves for 3C 120. For clarity, error bars have been omitted from the short term light curve. Average errors were 0.21 counts/sec.

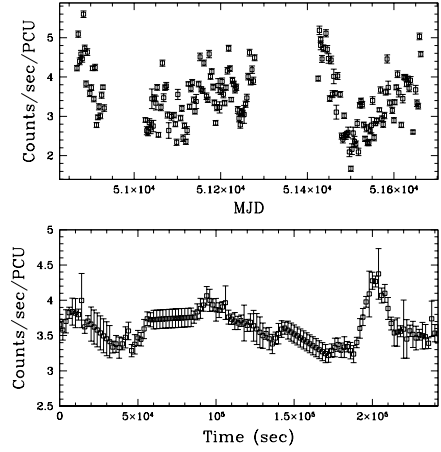

**FIGURE 2.** Light curves for Akn 120. The long-term light curve does not have the gaps due to sun passage interpolated. The short-term light curve covers MJD 51162–51165.

## X-RAY POWER DENSITY SPECTRA

Power spectra were computed using a discrete Fourier transform algorithm. No window function was applied to the data beforehand. The power spectra were normalized to units of squared fractional variability (rms$^2$/Hz). The individual PDSs for each light curve were binned geometrically every factor of 2 and then combined [5]. The binned power spectra for 3C 120 and Akn 120 are shown in Figure 3 and Figure 4, respectively.

Each PDS was then fit with a single power law, of the form $P \propto f^\alpha$, where $P$ is variability power, $f$ is temporal frequency, and $\alpha$ is the power law index. This gave an unacceptable fit for both objects, with $\chi^2$ 287 (10 degrees of freedom) for Akn 120 and $\chi^2$ 52 (16 degrees of freedom) for 3C 120.

Next we fit a broken power law of the form $P \propto f^{\alpha_1}$ below a break frequency $f_b$, and $P \propto f^{\alpha_2}$ above. This yielded a much better fit for both objects. For 3C 120, we found $\alpha_1$ 1.04, $\alpha_2$ 1.69, with $f_b$ $1.29 \times 10^{-6}$ Hz ($\chi^2$ 22.82 for 14 degrees of freedom). For Akn 120, we found $\alpha_1$ 1.10, $\alpha_2$ 1.83, with $f_b$ $1.84 \times 10^{-6}$ Hz ($\chi^2$ 20.64 for 8 degrees of freedom). The break timescales occur at roughly 9 and 6 days for 3C 120 and Akn 120, respectively.

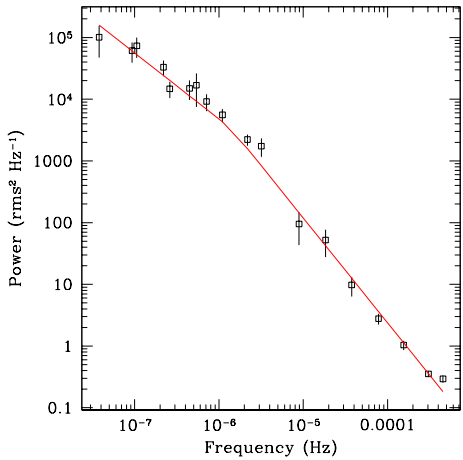

**FIGURE 3.** The PDS of 3C 120, with best-fit broken power law.

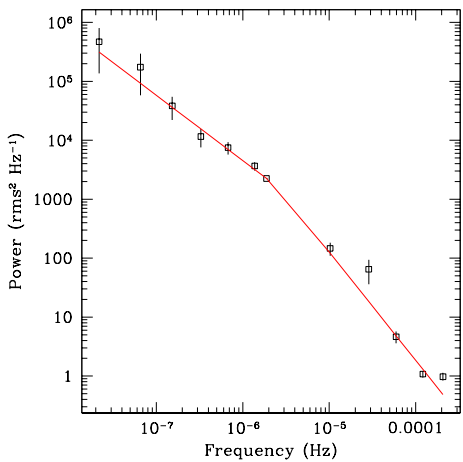

**FIGURE 4.** The PDS of Akn 120, with best-fit broken power law.

## OPTICAL VARIABILITY

To better constrain the relationship between optical and X-ray variability in AGN, we also obtained simultaneous optical data during our monitoring campaign of Akn 120. R and V filter observations were taken by a number of groups during the 2 years of X-ray monitoring, as well as intensive optical observations during the 3-day RXTE

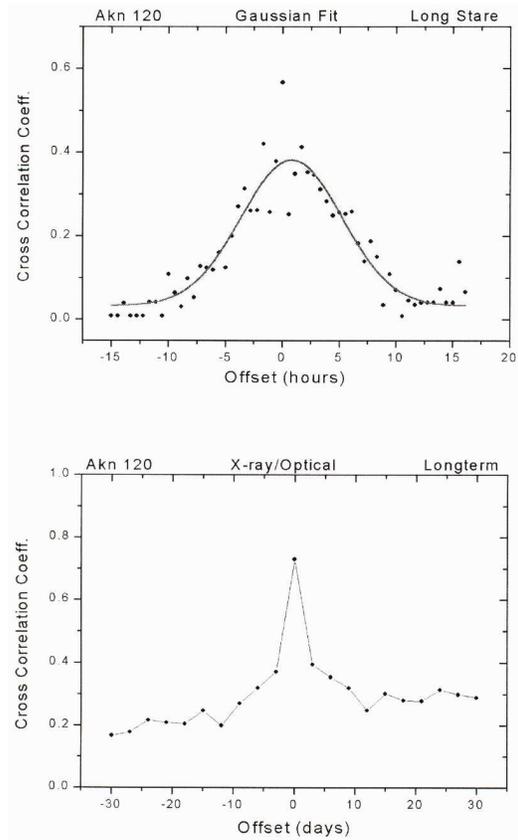

**FIGURE 5.** Cross-correlation functions for short (top) and long (bottom) timescale data

observation.

Cross-correlation coefficients were calculated for both the short and long timescale light curves. The cross-correlation was done using the MAKECC procedure for IDL, written by Gies, Penny, & Thaller. Figure 5 shows the cross-correlation functions for each data set.

The long-term data set shows a strong correlation coefficient of 0.75 at zero lag. This shows a significant relationship between X-ray and optical variability on a timescale less than the sampling frequency of 3 days. A cross-correlation of the optical and X-ray data for the intensive stare produces a noisy cross-correlation function. Fitting the result with a Gaussian function produces a coefficient of 0.38, with an offset of 0.79 hours, about 1.5 times the sampling rate. These results suggest that the variations are not strongly correlated on short timescales.

## ESTIMATION OF MASS

Galactic black hole binaries such as Cygnus X-1 typically exhibit two modes of X-ray variability: a low flux state with a hard spectrum, and a high flux state with a much softer spectrum [6]. These states also have different power spectra: the low/hard state exhibits a power law slope of 0 which breaks to –1 at 0.2 Hz, and then again to –2 at 4 Hz, while the high/soft state has only a single break, from a slope of –1 to –2 near 10-20 Hz [7, 8]. Most previous efforts have compared the PDSs of AGN to the low/hard state of Cygnus X-1, although recent efforts show that some galaxies may be best fitted by a high/soft state analogy [9].

Assuming that break frequency scales linearly with black hole mass, we can use the PDS of Cygnus X-1 to estimate the masses of the black holes in 3C 120 and Akn 120. For 3C 120, using the high frequency break of the low/hard state, the ratio of masses is $3.1 \times 10^6$. Previous efforts have placed the mass of the compact object in Cygnus X-1 at $10\,M_\odot$ [6], therefore we can estimate the mass of 3C 120 as $3.1 \times 10^7\,M_\odot$. This is remarkably close to the mass obtained from reverberation mapping, $3.0^{2.0}_{-1.5} \times 10^7\,M_\odot$ [10].

Using a time-averaged spectrum from the intensive 9-day stare, we find a 2–10 keV luminosity of $2.17 \times 10^{44}$ erg s$^{-1}$ for 3C 120. This implies a fractional Eddington accretion rate of $\dot{M}/\dot{M}_{Edd}$ 0.07. The low/hard state of Cygnus X-1 is thought to have a similarly low accretion rate.

For Akn 120, the picture is not as clear. If we scale the PDS break frequency with the high-frequency break from the low/hard state PDS of Cygnus X-1, we find a mass of $2 \times 10^7\,M_\odot$. This is roughly an order of magnitude less than the reverberation mapped mass of $1.86 \times 10^8\,M_\odot$. Because of this discrepancy, it is worth looking at whether the PDS of Akn 120 is more similar to the high/soft state PDS of Cygnus X-1.

Using a method to similar to 3C 120, we find a fractional Eddington accretion rate of $\sim 20\%$. This is significantly higher than what is seen in most AGN, including 3C 120. If we now scale the break seen in the PDS with the break at $\sim 15$ Hz seen in the high/soft state PDS of Cygnus X-1, we find a mass ratio of $8.15 \times 10^6$, implying a mass of $\sim 8 \times 10^7\,M_\odot$ for Akn 120. Although still too small, the fit is much better than the order of magnitude difference seen with the low/hard state PDS. We feel the high/soft state of Cygnus X-1 provides the best model for the high accretion rate and PDS of Akn 120.

## SUMMARY AND DISCUSSION

Using several years of RXTE monitoring, we have presented the PDSs for both 3C 120 and Akn 120. The PDSs for both objects show breaks from a slope of roughly $-1$ to $-2$, at timescales of $\sim 9$ and $\sim 6.3$ days for 3C 120 and Akn 120, respectively. The PDS of 3C 120 agrees very well with the low/hard state of the PDS for Cygnus X-1, allowing us to estimate the mass of the central compact object as $3.1 \times 10^7\,M_\odot$. This is nearly identical to the value obtained with reverberation mapping.

The PDS of Akn 120 agrees much better with the high/soft state of the PDS for Cygnus X-1, giving a mass estimate of $8 \times 10^7\,M_\odot$. This is slightly less than the reverberation mapped value of $\sim 2 \times 10^8\,M_\odot$. Cross-correlation with optical data shows a strong correlation with zero lag for the long term data, and a weaker correlation at shorter timescales, with optical variations lagging the X-rays by slightly less than an hour. However, more data are needed to strengthen any possible correlation at short timescales.

The radio emission and superluminal jet of 3C 120 provide insight into the nature of X-ray variability within these objects. The excellent correspondence between the PDS of 3C 120 and galactic black hole candidates such as Cygnus X-1, as well as the similarity to the PDSs of other, radio-quiet, AGN shows evidence that the mechanism of X-ray variability is the very similar for each class of object.

## ACKNOWLEDGMENTS

KM and HRM have been supported in part by an award from GSU's RPE Fund to PEGA, and by grants from the Research Corporation and NASA (NAGW-4397). APM has been supported by NASA grant NAG5-13074.